\def\simlt{\lower.5ex\hbox{$\; \buildrel < \over \sim \;$}}
\def\simgt{\lower.5ex\hbox{$\; \buildrel > \over \sim \;$}}
\RequirePackage{lineno} 
 \documentclass[preprint2]{emulateapj}

\slugcomment{Accepted to the Astrophysical Journal }
\shorttitle{Detection of Molecules}
\shortauthors{Swain et al.}

\begin{document}

\title{On the Detection of Molecules in the Atmosphere of HD189733b using HST NICMOS Transmission Spectroscopy}

\author{Mark R. Swain}
\affil{Jet Propulsion Laboratory, California Institute of Technology, 4800 Oak Grove Drive, Pasadena, CA 91109, USA}
\author{Michael R. Line}
\affil{Department of Astronomy \& Astrophysics, University of California-Santa Cruz, Santa Cruz, CA 95064}
\author{Pieter Deroo}
\affil{Jet Propulsion Laboratory, California Institute of Technology, 4800 Oak Grove Drive, Pasadena, CA 91109, USA}

\email{Mark.R.Swain@jpl.nasa.gov}

\begin{abstract}
The HST/NICMOS transmission spectrum measurements of HD 189733b that suggest the detection of methane (CH$_{4}$) in an exoplanet atmosphere have been a source of significant controversy. With what is probably the best analyzed exoplanet spectroscopy data set to date, different teams, using different methods, have claimed evidence both contradicting and supporting the original findings. Here, we report results from a uniform spectral retrieval analysis of the three, independent, published spectra together with null hypothesis testing.  Based on Bayesian model comparison, we find that two of the three spectra show strong evidence ($\geq$ 3.6$\sigma$) for the detection of molecular features mainly due to water and methane while the third is consistent with a weak molecular detection at the 2.2$\sigma$ level.  We interpret the agreement in the spectral modulation established by previous authors and the atmospheric retrieval results presented here, as a confirmation of the original detection of molecular absorbers in the atmosphere of HD 189733b. 
\end{abstract}

\keywords{planetary systems --- planets and satellites: atmospheres 
 --- radiative transfer--methods: data analysis--planets and satellites: individual(HD 189733b)}

\section{Introduction}

The announcement of the likely detection of methane in an exoplanet
atmosphere was made using a Hubble/NICMOS near-infrared transmission
spectrum of the hot-Jupiter HD 189733b by Swain, Vasisht, \& Tinetti
(2008; hereafter SVT08).  The same measurements, and those of
Grillmair et al. (2008), also provided the spectroscopic confirmation
of the presence of water, which had been previously identified using
Spitzer mid-infrared photometry (Tinetti et al. 2007).  The Hubble
spectra of HD 189733b initiated extensive efforts in the community to
characterize exoplanet atmospheres by searching for molecular features
using Hubble near-infrared spectroscopy measurements of both primary
eclipse (transit) and secondary eclipse (occultation) events; today,
infrared spectroscopic characterization of transiting exoplanet
atmospheres with Hubble is a robust field involving multiple teams,
hundreds of Hubble orbits, and published spectra for 11 planets to
date (HD 189733b, HD 209458b, GJ 436b, XO-1b, XO-2b, GJ 1214b,
WASP-12b, HAT-P-1b, HAT-P-12b, WASP-17b, WASP-19b - see Swain et
al. 2008, Pont et al. 2009, Swain et al. 2009a, Swain et al. 2009b,
Tinetti et al. 2010, Crouzet et al. 2012, Berta et al. 2012, Swain et
al. 2013, Huitson et al. 2013, Wakeford et al. 2013, Deming et
al. 2013, Line et al. 2013, Mandell et al. 2013, Kreidberg et
al. 2014).

As part of the growing interest in applying Hubble spectroscopy to
exoplanets, Gibson, Pont, \& Agrain (2011; hereafter GPA11) reanalyzed
the SVT08 data and produced three different transmission spectra,
based on three different models for the instrument systematics.  GPA11
proposed that the systematic errors in NICMOS were not amenable to
correction and that the results of SVT08 should not be considered
reliable.  Subsequently, Gibson et al. (2012; hereafter G12) applied a
new data reduction method and found âresults consistent with SVT08 but
with substantially larger errors.  On the basis of the larger errors,
G12 concluded that the detection of molecular features was unreliable.

In an attempt to resolve the debate, Waldmann et al. (2013; hereafter
W13), undertook a reanalysis of the SVT08 data using a completely
different approach from either SVT08 or GPA12.  This analysis resulted
in a spectrum consistent with the SVT08 and GPA12 results (see
Figure \ref{fig:figure1}).  W13 concluded that the agreement between
these three spectra is strong evidence for the stability of the
result.  W13 found measurement uncertainties 30\% larger than SVT08
and noted that the method they (W13) used, which does not make use of
any prior knowledge of the instrument, generates larger uncertainties
than an approach based on an instrument model used by SVT08.

Although the G12 claim that the expected signal is typically orders of
magnitude smaller than the instrumental systematics is demonstrably
incorrect, (see Figure \ref{fig:figure1} reproduced from SVT08), the
absence of a quantitative analysis of the constraints provided by the
three spectra has fostered speculation.  Here we report a uniform
analysis to determine how the differences between the SVT08, G12, \&
W13 transmission spectra impact our knowledge of the presence of
molecular absorbers in the atmosphere of HD 189733b.

\section{Methods}
\begin{figure*}
  \centering
    \includegraphics[width=6in]{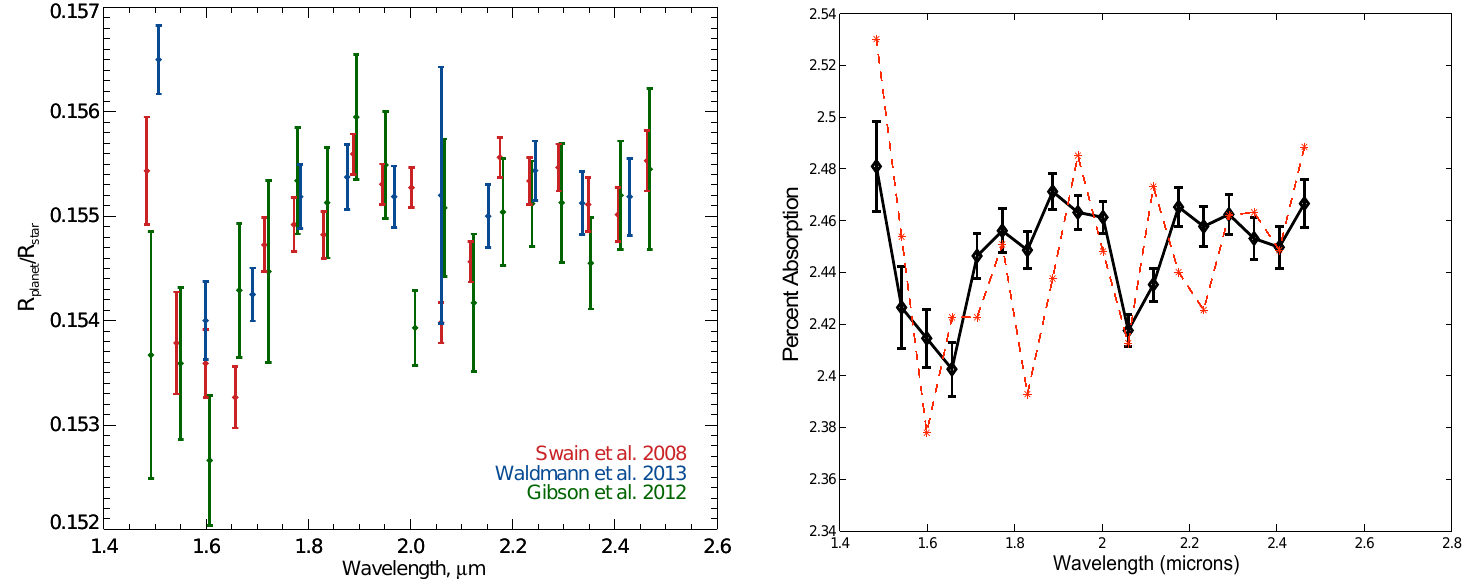}
     \caption{\label{fig:figure1}Corrected and uncorrected spectra: Left: Results for three independent reductions with completely different methods of the same NICMOS measurements of the transmission spectrum of HD 189733b.  The three results are similar in terms of shape, but there are differences in the uncertainties.  Right: The transmission spectrum before (red) and after (black) correction using an instrument model (Swain et al. 2008).  Differences between the red and black points for a given wavelength indicate the size of the instrument systematic errors which are relatively small.}
\end{figure*}

We explore the impact of each of the three spectra (see Figure \ref{fig:figure1}) on our knowledge of the composition of the atmosphere by asking the question: What is the detectability of molecular species in each of these spectra? We quantitatively answer this question using a Bayesian model comparison approach of the atmospheric retrieval results for each data set.  We used the Bayesian atmospheric retrieval suite, CHIMERA, described in detail in Line et al. (2013a) to perform the retrieval analysis.  CHIMERA uses three retrieval algorithms to determine the range of temperatures and abundances permitted by the data.  These three algorithms are optimal estimation (Rodgers 2000; Lee et al. 2012; Line et al. 2012), Markov chain Monte Carlo (MCMC-e.g., Madhusudhan et al. 2011; Benneke \& Seager 2012), and bootstrap Monte Carlo (Press et al. 1999).  Line et al. (2013a) showed that the three retrieval approaches produce temperature and molecular abundance uncertainty distributions that tend to disagree for spectra with sparse coverage and low signal-to-noise generally due to the non-Gaussian nature of the posterior.  The more widely accepted of the approaches in theses situations are the MCMC methods because of their ability to characterize non-Gaussian posterior probability distributions.  In this investigation we use both optimal estimation and MCMC for the model comparisons.

The forward model used in the retrievals computes the transmission spectra given the gas abundances and temperature profile.  The model divides the disk of the planet into annuli and computes the integrated slant optical depth and transmittance along each tangent height.  The wavelength-dependent eclipse depth is then computed by integrating the slant transmittance using equation 11 in Brown (2001).  The absorption cross-section database is described in Line et al. (2013a).  The code has been validated against those of Fortney et al. (2010) and Deming et al. (2013) (see Line et al. 2013b for the validation).  

Our objective, for all three spectra, is the retrieval of the constant-with-altitude volume mixing ratios for H$_{2}$O, CH$_{4}$, CO, and CO$_{2}$; assumptions and retrieval parameters are applied in a uniform way for all three spectra.  We assume an isothermal atmosphere, a valid assumption for transmission spectra due to the relative insensitivity of the transmission spectra to changes in this temperature range and to the lack of justification for more complicated profiles.  This single temperature parameter controls the scale height, and hence the absorption feature amplitudes. Additionally, we fix the planet radius and adjust the reference pressure level at which this radius is defined; this slides the overall transmission spectrum up or down.  Thus the retrieval is for a total of six parameters: the four gases, temperature, and reference pressure.  The mean molecular weight of the atmosphere is self consistently determined using the mole fractions of the four retrieved species and assuming the remainder of the atmosphere is a cosmic H$_{2}$/He mixture.  There may be other optically inactive filler gases such as N$_{2}$ or noble gases; however, given the mass and radius measurements of HD189733b, H$_{2}$/He are the species that would most contribute to the mean molecular weight.  We have not considered the role of clouds in these spectra.  We assume flat priors on each of the parameters for the MCMC retrieval.  The details of these priors can be found in Line et al. (2013a).

\section{Results}

\begin{figure}
  \centering
    \includegraphics[width=0.45\textwidth]{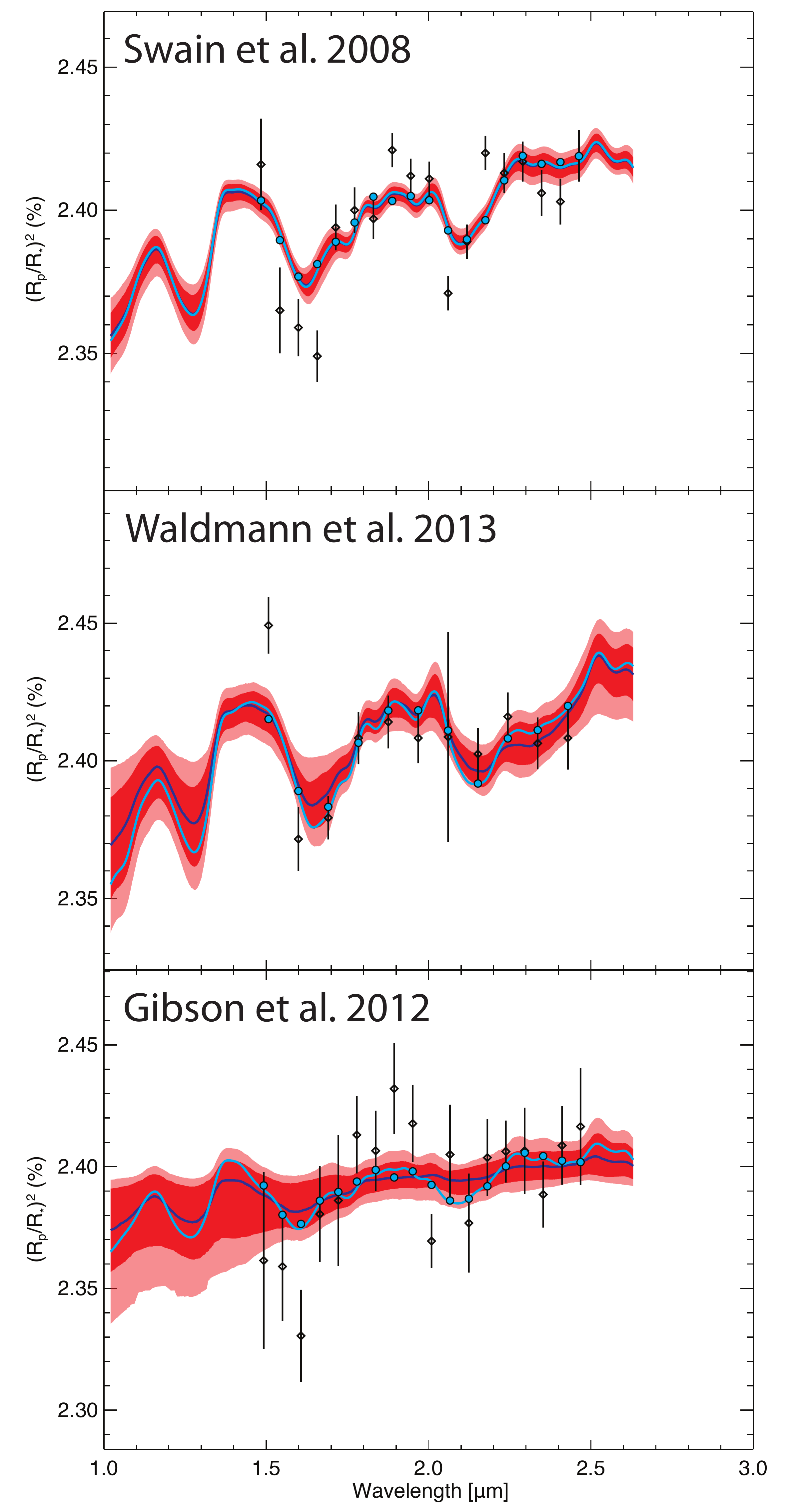}
     \caption{\label{fig:figure2} Retrieval fits to the data.  The MCMC retrieval produces many hundreds of thousands of spectra.  We summarize the spread in the spectra with a median (dark blue), and 68\% (dark red) and 95\% (light red) confidence intervals.  The light blue curve is the best fit.  The light blue circles are the best fit model binned to the data. For each data set, the model predictions for 1.1 to 1.5 $\mu$m are included to facilitate comparison with future WFC3 IR grism observations.}
\end{figure}

\begin{figure*}
  \centering
    \includegraphics[width=0.85\textwidth]{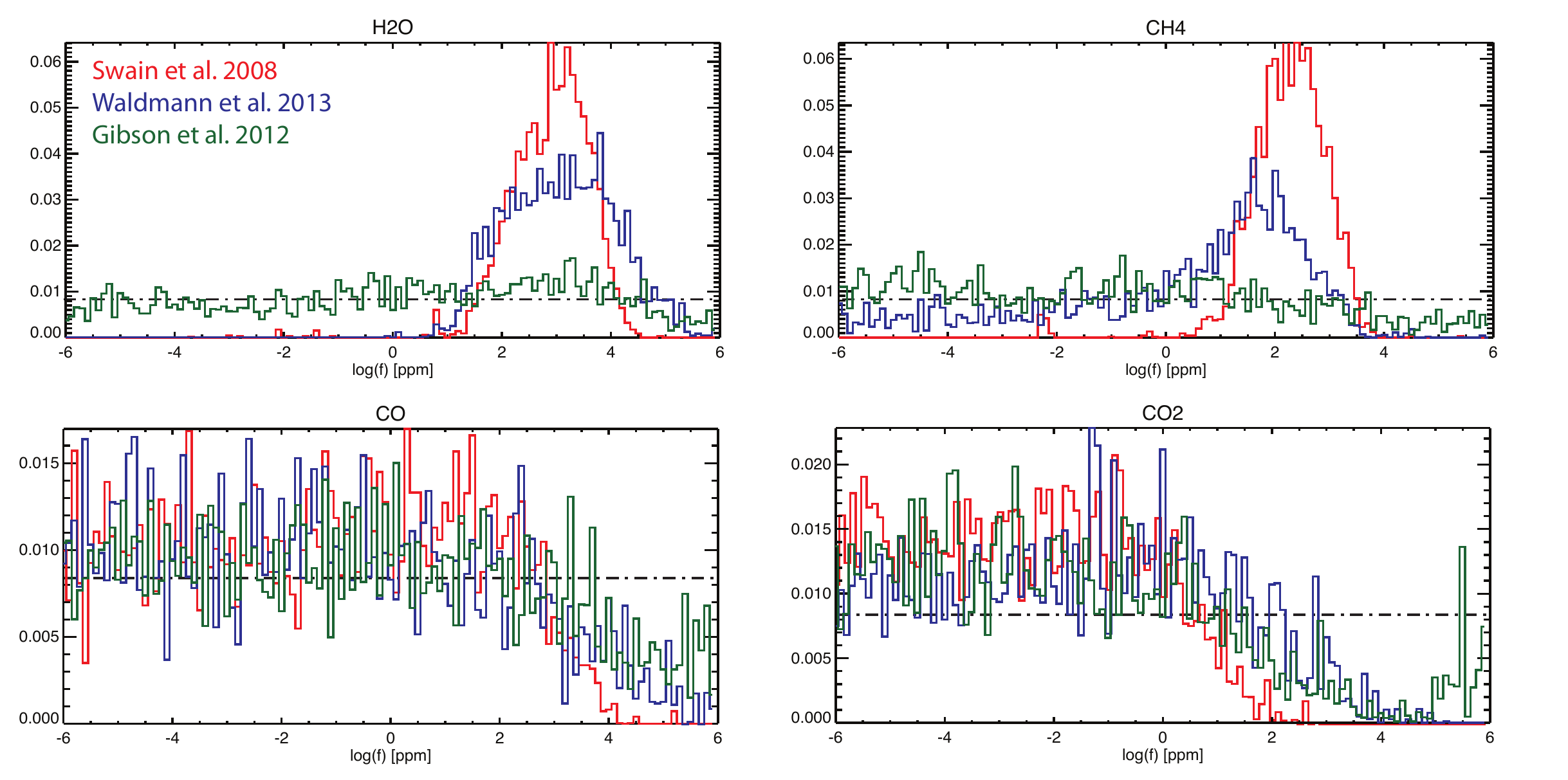}
     \caption{\label{fig:figure3}Marginalized gas posteriors resulting from each data set. The x-axis is the log of the gas mixing ratios in parts-per-million (ppm).  The horizontal blue line in each is the flat prior used in the MCMC retrieval.}
\end{figure*}

Figures \ref{fig:figure2}  and \ref{fig:figure3} summarize the gas abundance retrieval results.  Figure \ref{fig:figure2} shows the fits to each of the three spectra.  Since the MCMC produces many hundreds of thousands of fits, rather than show all of them, we summarize the fits by computing the median of all of the fits and the 68\% and 95\% spread in the fits (see Line et al. 2013a for details).  Generally, the spread in the fits roughly mimics the average data error bar size.  Figure \ref{fig:figure3} shows the marginalized posterior distributions for each gas in the form of histograms.  From this figure it is clear that the SVT08 spectrum produces the best constraints on the water and methane abundances.  The W13 spectrum produces similar constraints on water, and shows methane is likely present but provides less of a constraint on the abundance.  Finally, the G12 spectrum produces virtually no constraint on any of the gas abundances.  All three datasets fail to provide meaningful constraints on the CO and CO$_{2}$  abundances with only perhaps hinting at upper limits near mixing ratios of $\sim 10^{-1}$ and $\sim 10^{-2}$ respectively.  

In order to quantitatively determine the detectability of molecules within the spectra, we use two hypothesis-testing procedures.  The two hypotheses being tested are: This spectra suggests molecular absorptionâ and the null hypothesis: This model does not suggest any molecular absorption.   The first test is a Frequentists $\Delta \chi^{2}$ model comparison.  We compute the difference in $\chi^{2}$ between the best fit from the full model with all of the gases (six total parameters) and the best-fit null model without any of the gases (two total parameters).  The best fit atmospheric state is determined using optimal estimation.  This $\Delta \chi^{2}$ and the change in degrees of freedom (four) can be used to compute a $p$-value, or the probability of obtaining a larger delta chi-squared for repeated sets of measurements.  This $p$-value can then be converted into a significance level (e.g., Gregory 2005).  Table \ref{tab:table1} shows the results.  If we take 3.6$\sigma$ to be the criterion for a significant detection (Trotta 2008), then from this test we find that only the SVT08 and W13 data allow for statistically significant molecular detections.  Although consistent with molecular absorption, the G12 data do not constitute a molecular detection as measured by the $\Delta \chi^{2}$ test.

\begin{deluxetable}{cccc}
\tablecolumns{4} 
\tablewidth{0pt}  
\tablecaption{\label{tab:table1}$\Delta \chi^{2}$ test results.  The $\chi^{2}$ values are from the best fits for each scenario and data set.  The null model is the gas free model. The change in degrees of freedom is four.}
\tablehead{ &\colhead{SVT08} & \colhead{W13} & \colhead{G12} }
\startdata
With Gases $\chi^{2}$ &   66.51  &   18.32 & 23.02  \\ \hline
Null $\chi^{2}$       &  100.25  &   40.87 & 25.68  \\ \hline
$\Delta \chi^{2}$     &   33.74  &   22.55 &  2.66  \\
$p$-value & 3.98$\times 10^{-7}$ & 7.15$\times 10^{-5}$ & 0.176 \\
\bf{Detection Level ($\sigma$)}  & \bf{5.1} & \bf{4.0} & \bf{1.4} 
\enddata
\end{deluxetable}
The second hypothesis-testing procedure is considered more rigorous as it relies on the Bayesian evidence (also known as the marginal likelihood) resulting from the MCMC retrievals (see e.g., Benneke \& Seager 2013) and it is these findings we highlight in the abstract. The Bayesian evidence is a multidimensional integral over the volume of phase space explored by the MCMC. The computation of this integral is non-trivial and there are numerous approaches to compute the evidence such as the harmonic mean (Newton and Raftery 1994), Laplace approximation (e.g., Kass and Raftery 1995), nested sampling (Skilling 2004), and others.  Each approach has its advantages and pitfalls.  We choose to use the Numerical Lebesgue Algorithm (NLA) approach (Weinberg 2012), which is a variant of the harmonic mean approximation but solves the problem of the large truncation error.  This approach is straightforward to implement and can be applied to the MCMC chains generated by CHIMERA.

We compute the evidence for the full model and the null (gas free)
model for each of the three data sets.  The ratio of the evidences
produces a Bayes factor.  This Bayes factor can then be converted to a
$p$-value and confidence interval (see e.g., Sellke et al. 2001;
Trotta 2008; Benneke \& Seager 2013).  Table \ref{tab:table2} shows
the results from this hypothesis testing procedure.  Again, consistent
with the $\Delta \chi^{2}$ test, we find that the SVT08 data results
in the largest molecular detection followed by the W13 and G12 data.

We have shown through both a Frequentist and Bayesian model comparison
exercise that the SVT08 data provides the strongest evidence for
molecular detection followed closely by the W13 data.  In contrast,
the G12 data are consistent with the presence of molecular features
and, at best, constitute a weak detection as measured by a Bayesian
model comparison, but do not constrain abundances for any of the
molecular species.

In displaying the results for the range of models fit to each
spectrum, we have included the model prediction for the 1.1 to 1.5
$\mu$m wavelength regions.  The models fit to the SVT08 and W13 data
predict the 1.35 $\mu$m water opacity feature, whereas the model fit
to the G12 data can be consistent with a wide range of possibilities,
ranging from flat to significant spectral modulation. The model
predictions in the 1.1 to 1.5 $\mu$m spectral region are displayed to
facilitate potential comparison of these models with WFC3 IR grism
observations of the HD 189733b transmission spectrum.  Extending the
wavelength range of the transmission spectrum is highly desirable;
however, there are two critical caveats to consider.  First, the
model results here only provide a prediction and should be updated
with a model fitting all the data if and when WFC3 results for this
object become available.  Second the WFC3 IR grism and NICMOS
measurements are taken $\sim$6 years apart, and the amount of haze,
inferred from visible measurements (e.g. Sing et al. 2011), in the
planet's atmosphere could have changed.  Notwithstanding these caveats, the models fit
to the SVT08 and W13 data predict a water absorption feature of
$\sim$300 to 400 ppm in the WFC3 IR grism passband.

\begin{deluxetable}{cccc}
\tablecolumns{4} 
\tablewidth{0pt}  
\tablecaption{\label{tab:table2}Bayesian model comparison resulting from the MCMC retrievals.  $Z_{0}$ is the evidence from the full model, which includes all of the gases.  $Z$ is the evidence computed from the null model.  $B$ is the Bayes factor which is the ratio of $Z_{0}$ to $Z$.}
\tablehead{ &\colhead{SVT08} & \colhead{W13} & \colhead{G12} }
\startdata
With Gases $ln(Z_{0})$ &  127.22  &   87.96 & 133.77  \\ \hline
Null $ln(Z)$          &  110.95  &   77.63 & 132.45  \\ \hline
$ln(B)$               &   16.27  &   10.33 &   1.32  \\ 
$p$-value & 1.56$\times 10^{-9}$ & 8.60$\times 10^{-7}$ & 0.027 \\
\bf{Detection Level ($\sigma$)}  & \bf{6.0} & \bf{4.9} & \bf{2.2} 
\enddata
\end{deluxetable}

\section{Discussion}

While the retrieval results for all three spectra are consistent with the presence of water and methane, there are significant differences in the degree of constraint the spectra provide on the gas abundances.  The three spectra represent very different approaches, undertaken by different practitioners, to determining the exoplanets spectrum.  One might rightly ask, which of these three spectra should be used in studies of HD 189733b?  We recommend use of the SVT08 result for three reasons.  First, the relative similarity for the significance of molecular detection in SVT08 and W13 suggests the G12 spectrum represents a less optimal treatment of the data.  Second, the W13 approach, as clearly stated in their paper, does not represent the optimal method for estimating the spectrum and is expected to provide larger measurement uncertainties than the SVT08 method.  Third, the extensive level of due diligence, outlined below, that was applied to the SVT08 data.

\begin{figure}
  \centering
    \includegraphics[width=0.45\textwidth]{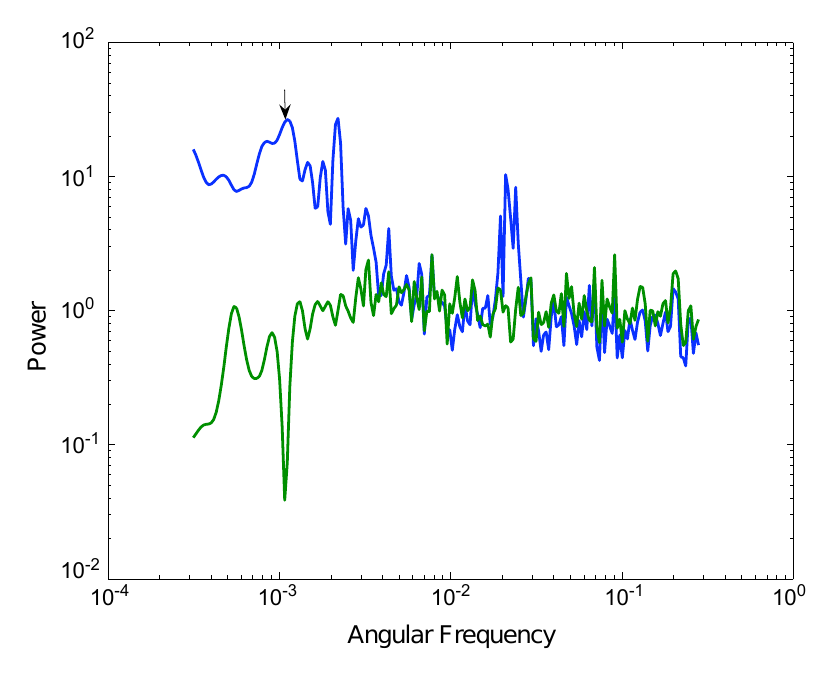}
     \caption{\label{fig:figure4}The periodogram for measured spectral flux showing significant red noise (blue plot) that is removed by the instrument model (green plot).   The arrow indicates the orbital period of Hubble.  This illustrates one of the kinds of tests done to validate the instrument model used for correcting the data in the SVT08 analysis.}
\end{figure}

The approach used in the SVT08 analysis, based on experience with instrumentation (Swain et al. 1998, 1999, 2003, 2004; Vasisht et al. 1998, 2003, 2004, 2006), was to assume systematic errors were present, and to exhaustively search the data to identify and remove these errors through modeling the instrument performance in terms of basic, measurable, instrument properties. The SVT08 team used the image and header data to construct ancillary data products that measured basic instrument characteristics such as pointing, focus, grism rotation, observatory orbital phase, and focal plane array temperature.  Using the out-of-eclipse spectrophotometric time series, with the assumption that temporal changes in the measured spectral photometric flux were due to linear changes in these instrument parameters, a model for the measured flux was constructed.  The linearity assumption was explicitly tested and, for one parameter (orbital phase), it was found that inclusion of a dependence on the square of this parameter decreased variance in the model-data residuals.  Periodograms confirmed that the modeling process effectively removed the temporal correlations from the spectral photometric time series (see Figure \ref{fig:figure4}).  A small amount of wavelength-correlated noise (~30\% of the random noise) was identified and removed by subtraction of an optimally weighted channel average for each sample. The instrument model plus an astrophysical model incorporating limb darkening was then applied to the data. The robustness of the spectral eclipse depth estimate was verified by extensive data removal and refitting as well as investigating the possible affects of star spots.  The magnitude of the corrections applied by the instrument model was determined (see Figure \ref{fig:figure1}), and consistency of the model-corrected broadband eclipse depth with nonmodel-corrected broad band eclipse depth was confirmed.  An extensive description of the calibration process summarized here appeared in the SVT08 supplementary material and we refer the reader to that source for further information.

\section{Conclusions}

Two versions of the calibrated spectra suggest the presence of molecular absorbers at high confidence.  All three methods show the presence of a combination of water and/or methane is probable.  The fact that three different data reduction methods have produced similar spectra giving similar spectral retrieval results in two cases, and consistent results in all three, is a remarkable validation of both this data set and the NICMOS instrument in general.  This level of independent results confirmation is unique in exoplanet data analysis and is a tribute to the hard work and dedication of all the teams involved.  Based on this collective effort, we can resolve an ongoing debate and state with a high degree of confidence that the NICMOS measurements show the presence of molecular absorbers in the atmosphere of HD 189733b.

\section{Acknowledgements}
The authors thank Gautam Vasisht for comments on this manuscript. The research described in this publication was carried out in part at the Jet Propulsion Laboratory, California Institute of Technology, under a contract with the National Aeronautics and Space Administration. Copyright 2013. All rights reserved.

\end{document}